%Paper: cond-mat/9503110
%From: Denjoe O'Connor <denjoe@stp.dias.ie>
%Date: Sun, 19 Mar 1995 19:01:40 +0000 (GMT)

%The Specific Heat of a Ferromagnetic Film.
\font\bigg=cmbx10 at 17.3 truept
\font\big=cmbx10 at 12 truept
\def\cl{\centerline} %\def\ex{{\rm e}}

%%%%%%%%%%%%%%%%%%%%%%%%%%%%%%%%%
%%macros from envfrin
%%%%%%%%%%%%%%%%%%%%%%%%%%%%%%%%%

\font\twelverm=cmr10 scaled 1200    \font\twelvei=cmmi10 scaled 1200
\font\twelvesy=cmsy10 scaled 1200   \font\twelveex=cmex10 scaled 1200
\font\twelvebf=cmbx10 scaled 1200   \font\twelvesl=cmsl10 scaled 1200
\font\twelvett=cmtt10 scaled 1200   \font\twelveit=cmti10 scaled 1200

\skewchar\twelvei='177   \skewchar\twelvesy='60

%  Define \...point macros to change fonts and spacings consistently

\def\twelvepoint{\normalbaselineskip=12.4pt
  \abovedisplayskip 12.4pt plus 3pt minus 9pt
  \belowdisplayskip 12.4pt plus 3pt minus 9pt
  \abovedisplayshortskip 0pt plus 3pt
  \belowdisplayshortskip 7.2pt plus 3pt minus 4pt
  \smallskipamount=3.6pt plus1.2pt minus1.2pt
  \medskipamount=7.2pt plus2.4pt minus2.4pt
  \bigskipamount=14.4pt plus4.8pt minus4.8pt
  \def\rm{\fam0\twelverm}          \def\it{\fam\itfam\twelveit}%
  \def\sl{\fam\slfam\twelvesl}     \def\bf{\fam\bffam\twelvebf}%
  \def\mit{\fam 1}                 \def\cal{\fam 2}%
  \def\tt{\twelvett}
  \textfont0=\twelverm   \scriptfont0=\tenrm   \scriptscriptfont0=\sevenrm
  \textfont1=\twelvei    \scriptfont1=\teni    \scriptscriptfont1=\seveni
  \textfont2=\twelvesy   \scriptfont2=\tensy   \scriptscriptfont2=\sevensy
  \textfont3=\twelveex   \scriptfont3=\twelveex  \scriptscriptfont3=\twelveex
  \textfont\itfam=\twelveit
  \textfont\slfam=\twelvesl
  \textfont\bffam=\twelvebf \scriptfont\bffam=\tenbf
  \scriptscriptfont\bffam=\sevenbf
  \normalbaselines\rm}

%       tenpoint

%%
%%      Various internal macros
%%

\def\beginparmode{\endmode
  \begingroup \def\endmode{\par\endgroup}}
\let\endmode=\par
{\obeylines\gdef\
{}}
\def\singlespace{\baselineskip=\normalbaselineskip}

\def\doublespace{\baselineskip=\normalbaselineskip \multiply\baselineskip by 2}

\newcount\firstpageno
\firstpageno=2
\footline={\ifnum\pageno<\firstpageno{\hfil}\else{\hfil\twelverm\folio\hfil}\fi}
\let\rawfootnote=\footnote              % We must set the footnote style
\def\footnote#1#2{{\rm\singlespace\parindent=0pt\rawfootnote{#1}{#2}}}
\def\raggedcenter{\leftskip=4em plus 12em \rightskip=\leftskip
  \parindent=0pt \parfillskip=0pt \spaceskip=.3333em \xspaceskip=.5em
  \pretolerance=9999 \tolerance=9999
  \hyphenpenalty=9999 \exhyphenpenalty=9999 }
\def\dateline{\rightline{\ifcase\month\or
  January\or February\or March\or April\or May\or June\or
  July\or August\or September\or October\or November\or December\fi
  \space\number\year}}
\def\received{\vskip 3pt plus 0.2fill
 \centerline{\sl (Received\space\ifcase\month\or
  January\or February\or March\or April\or May\or June\or
  July\or August\or September\or October\or November\or December\fi
  \qquad, \number\year)}}

%%
%%      Page layout, margins, font and spacing (feel free to change)
%%

\hsize=6.5truein
\hoffset=.1truein
\vsize=8.9truein
\voffset=.05truein
\parskip=\medskipamount
\twelvepoint            % selects twelvepoint fonts (cf. \tenpoint)
\doublespace            % selects double spacing for main part of paper (cf.
                        %       \singlespace, \oneandahalfspace)
\overfullrule=0pt       % delete the nasty little black boxes for overfull box

\def\preprintno#1{
 \rightline{\rm #1}}    % Preprint number at upper right of title page

\def\head#1{                    % Head;  NOTE enclose the text in {}
  \filbreak\vskip 0.5truein     %  e.g., \head{I. Introduction}
  {\immediate\write16{#1}
   \raggedcenter \uppercase{#1}\par}
   \nobreak\vskip 0.25truein\nobreak}

\def\references                 % Begin references -- basic format is Phys Rev
  {\head{References}            % I.e., volume, page, year (space after
%%commas).
   \beginparmode
   \frenchspacing \parindent=0pt \leftskip=1truecm
   \parskip=8pt plus 3pt \everypar{\hangindent=\parindent}}

\def\frac#1#2{{\textstyle{#1 \over #2}}}
\def\square{\kern1pt\vbox{\hrule height 1.2pt\hbox{\vrule width 1.2pt\hskip 3pt
   \vbox{\vskip 6pt}\hskip 3pt\vrule width 0.6pt}\hrule height 0.6pt}\kern1pt}
%%%%%%%%%%%%%%%%%%%%%%%%%%%%%%%%%%%%%%%%%%%%%%%%%%%%%%%%%%%%%%%%%%%%%
%nashmac
\newcount\pagenumber
\newcount\sectionnumber
\newcount\appendixnumber
\newcount\equationnumber

\newcount\citationnumber
\global\citationnumber=1

\def\ifundefined#1{\expandafter\ifx\csname#1\endcsname\relax}
\def\cite#1{\ifundefined{#1} {\bf ?.?}\message{#1 not yet defined,}
\else \csname#1\endcsname \fi}

\def\docref#1{\ifundefined{#1} {\bf ?.?}\message{#1 not yet defined,}
\else \csname#1\endcsname \fi}

\def\article{
\def\eqlabel##1{\edef##1{\sectionlabel.\the\equationnumber}}
\def\seclabel##1{\edef##1{\sectionlabel}}
\def\citelabel##1{\edef##1{\the\citationnumber}{\global\advance\citationnumber
by1}}
}

\def\appendixlabel{\ifcase\appendixnumber\or A\or B\or C\or D\or E\or
F\or G\or H\or I\or J\or K\or L\or M\or N\or O\or P\or Q\or R\or S\or
T\or U\or V\or W\or X\or Y\or Z\fi}

\def\sectionlabel{\ifnum\appendixnumber>0 \appendixlabel
\else\the\sectionnumber\fi}

\def\beginsection #1
 {{\global\subsecnumber=1\global\appendixnumber=0\global\advance\sectionnumber
by1}\equationnumber=1
\par\vskip 0.8\baselineskip plus 0.8\baselineskip
 minus 0.8\baselineskip
\noindent $\S$ {\bf \sectionlabel. #1}
\par\penalty 10000\vskip 0.6\baselineskip plus 0.8\baselineskip
minus 0.6\baselineskip \noindent}

\newcount\subsecnumber
\global\subsecnumber=1

\def\subsec #1 {\bf\par\vskip8truept  minus 8truept
\noindent \ifnum\appendixnumber=0 $\S\S\;$\else\fi
$\bf\sectionlabel.\the\subsecnumber$ #1
\global\advance\subsecnumber by1
\rm\par\penalty 10000\vskip6truept  minus 6truept\noindent}

\def\beginappendix #1
{{\global\subsecnumber=1\global\advance\appendixnumber
by1}\equationnumber=1\par
\vskip 0.8\baselineskip plus 0.8\baselineskip
 minus 0.8\baselineskip
\noindent
{\bf Appendix \appendixlabel . #1}
\par\vskip 0.8\baselineskip plus 0.8\baselineskip
 minus 0.8\baselineskip
\noindent}

\def\no{\eqno({\rm\sectionlabel}
.\the\equationnumber){\global\advance\equationnumber by1}}

\def\ref #1{{\bf [#1]}}

\article
%%%%%%%%%%%%%%%%%%%%%%%%%%%%%%%%%%%%%%%%%%%%%%%%%%%%%%%%%%%%%%%%%%
%defsnew
\def\i{\infty}  %overwrites plain tex defn
\def\L{\Lambda} %overwrites plain tex defn
\def\l{\lambda} %overwrites plain tex defn
   %overwrites plain tex defn
    %overwrites plain tex defn
\def\o{\over}   %overwrites plain tex defn
   %overwrites plain tex defn

         \def\p{\partial}
\def\G{\Gamma}   \def\k{\kappa}
\def\ra{\rightarrow}         
\def\vft{\varphi^2}            
\def\e{\varepsilon}  
\def\gf{\gamma_{\varphi}}        

\def\vf{\varphi}
\def\r{\rho}
\def\gft{\gamma_{\vf^2}}  

\def\frac#1#2{{{#1}\over{#2}}}
\def\s{\scriptstyle}     \def\ss{\scriptscriptstyle}

\def\gl{\gamma_{\l}}

\def\nf{{(5N+22)\o9}}

\def\nt#1{{(N+2)\over{#1}}}

\def\io{\hbox{$\bigcirc\kern-2.3pt{\bf\cdot}$\kern 2.3pt}}
%tadpole
\def\bub{\hbox{${\bf\cdot}\kern-5pt\bigcirc\kern-5pt{\bf\cdot}$}}
%bubble
\def\itr{\hbox{${\bf\cdot}\kern-6pt\in\kern-2.5pt\ni\kern-6pt{\bf\cdot}\kern
2.5pt$}}
%setting sun
\def\itrp{\hbox{${\bf\cdot}\kern-6pt\in\kern-2.5pt\ni\kern-6pt{\bf\cdot}
\kern-10.5pt/$\kern 5pt}}
%wavefn ren
\def\if{\hbox{${\bf\cdot}\kern-3pt\langle\kern-2.4pt|\kern-2.2pt)$}}
%cone
   %1loop free energy
%mass insert
\def\itro{\hbox{${\bf\cdot}\kern-5pt\bigcirc\kern-5pt{\bf :}$}}
%fourdots
\def\bfour{\hbox{${\bf :}\kern-5pt\bigcirc\kern-5pt{\bf :}$}}

\def\Aif{A_{2(4-d)}\left(\if-{1\o2}\bub^2\right)}

\def\at{\nf A_{2(4-d)}\left(\if-{1\o2}\bub^2\right)+\nt{9}A_{2(4-d)}\itrp}

\def\tauxy#1#2{\tau r(x,y)+{{\omega(n_{1},n_{2},x,y)\o\k^2L^2}}}

\def\ef{_{\ss eff}}
%%%%%%%%%%%%%%%%%%%%%%%%%%%%%%%%%%%%%%%%%%%%%%%%%%%%%%%%%%%%%%%%
%%additional definitions:
\def\Got{\G^{(0,2)}}
\def\Gnl{\G^{(N,L)}}

%%additional definitions for specific heat paper

\def\inr{\int_1^{\rho}}

%The paper proper

\doublespace

\vskip -0.4truein
\preprintno{ICN-UNAM-94-08}
\preprintno{DIAS-STP-94-42}
\preprintno{HD-THEP-94-50}
%\dateline
\cl{\bigg{The Specific Heat of a Ferromagnetic Film}}
\vskip .5truecm
\cl{\bf F. Freire}
\cl{Inst. f\"ur Theor. Phys., Universit\"at Heidelberg}
\cl{Philosophenweg 16, 69120 Heidelberg, Germany}
\cl{\bf Denjoe O'Connor}
\cl{DIAS, 10 Burlington Road, Dublin 4, Ireland}
\cl{\bf C. R. Stephens}
\cl{DIAS, 10 Burlington Road, Dublin 4, Ireland}
\cl{and} \cl{Instituto de Ciencias Nucleares, UNAM,}
\cl{Circuito Exterior, A.\ Postal 70-543, M\'exico D.F. 04510.
\footnote{\dag}{Permanent address}}
\vskip 0.5truecm
\noindent{\bf Abstract:}\ \ We analyze the specific heat for the
$O(N)$ vector model on a $d$-dimensional film geometry
of thickness $L$ using ``environmentally friendly'' renormalization. We
consider
periodic, Dirichlet and antiperiodic boundary conditions, deriving expressions
for the specific heat and an effective specific heat exponent,
$\alpha\ef$. In the case
of $d=3$, for $N=1$, by matching to the exact exponent of the two dimensional
Ising model we capture the crossover for $\xi_L\ra\infty$ between power law
behaviour in the limit ${L\over\xi_L}\ra\infty$ and logarithmic behaviour in
the limit ${L\over\xi_L}\ra0$ for fixed $L$, where $\xi_L$ is the correlation
length
in the transverse dimensions.\hfill\break
PACS numbers: 64.60.Fr, 05.70.Jk, 68.35.Rh, 68.15.+e

\vfil \eject

\beginsection{Introduction}
Thermodynamic quantities generally depend on many details of the system,
and are therefore functions of a large number of variables,
however, in the critical regime this dependence drops to a smaller number. The
resulting functions are referred to as scaling functions.
Scaling functions generically describe a crossover, wherein the effective
degrees of freedom of a system can change dramatically as a function of scale.
Calculating such scaling functions in critical
phenomena is generally accepted to be much more difficult than calculating
critical exponents. From a renormalization group (RG) point of view one
can think of this as being due to the fact that to
calculate a critical exponent one only needs a local RG linearized around the
fixed
point of interest whereas, generally speaking, to calculate a scaling function
one needs a global, non-linear RG that is capable of encompassing more
than one fixed point. One of the chief difficulties in the latter is
developing a ``uniform'' approximation scheme that can describe the crossover
between two fixed points perturbatively. Conventional small parameters
such as $\e$ and $1/ N$ might be adequate for certain crossovers but not
others.

Crossovers are induced by some asymmetry parameter which often can be
fruitfully thought of as an ``environmental'' variable, such as temperature,
system size, magnetic field etc.
The formalism of ``environmentally friendly'' renormalization
\citelabel{\NucJphysa}\citelabel{\EnvfRG}
[\cite{NucJphysa},\cite{EnvfRG}] offers
a quite general approach to the solution of crossover problems and the
calculation of scaling functions. Given that the key idea behind the notion
of a crossover is the qualitatively changing nature of the effective degrees
of freedom as a function of ``scale'' and ``environment'' it implements a
renormalization which is capable of tracking the evolving effective degrees
of freedom in a perturbatively controllable manner.
However, it is based on reparametrization invariance, as in the original field
theoretic RG,
rather than Wilson/Kadanoff coarse graining.

The basic idea is that the relation of the bare couplings to the renormalized
ones, which can be used to describe, parametrically, a physical system,
can be thought of as a coordinate transformation in the space of these
couplings. In thinking of the renormalized couplings as new
``coordinates'' the conventional field theoretic RG simply expresses the
invariance of physical quantities under changes of coordinate system.
This coordinate invariance is an {\it exact} invariance of field theory.
When calculating a physical quantity perturbatively,
in spite of the fact that physics doesn't depend on coordinates,
the particular choice of coordinates can be quite crucial in obtaining a
reliable approximation scheme.
One can understand this clearly in the context of the crossover studied in
this paper --- dimensional crossover induced by finite size effects.

Reverting for the
moment to a coarse graining RG, if we thought of possible coarse grainings
in a $d$ dimensional ferromagnetic film of size $L$, one would find that
block spins of size $\xi\ll L$ were $d$ dimensional, whilst those of size
$\xi\gg L$ were $d-1$ dimensional. Thus this coarse graining procedure
reflects a crucial property of the ``environment'' of the system --- that it
is finite in one dimension. Block spinning is therefore an environmentally
friendly form of renormalization.  In the context of
reparametrization, an environmentally friendly renormalization is one
that yields a set of parameters that give a perturbatively reliable
description of the crossover. In the finite size context a necessary
condition for the reparametrization to be environmentally friendly is that
it be $L$ dependent.

In previous papers environmentally friendly renormalization has been used
to describe various physical quantities for particular crossovers of interest
\citelabel{\Procroys}\citelabel{\EqnofState}\citelabel{\PRL}
[\cite{Procroys}---\cite{PRL}].
In this paper we consider dimensional crossover of the specific heat
as it is one of the more readily accessible quantities from an experimental
point of view. In the context of films and experimental tests of finite size
scaling this was the first experimentally measured quantity
\citelabel{\Gasparini}[\cite{Gasparini}].

{}From a theoretical point of view the case of a totally finite geometry has
been successfully
investigated numerically \citelabel{\Binder}[\cite{Binder}] and analytically
for both periodic boundary conditions \citelabel{\RudGuoJas}
[\cite{RudGuoJas}] and Dirichlet boundary conditions
\citelabel{\HuhnDomb}[\cite{HuhnDomb}]. This case has recently
\citelabel{\ChenDohmEsser}[\cite{ChenDohmEsser}] been further developed
with particular emphasis paid to the problems created by the existence,
for an $n$ component order parameter, of
massless spin-waves (Goldstone modes).
In the case of film geometries some progress has been made
\citelabel{\KrechDietrich}[\cite{KrechDietrich}]
but no theoretical work has been able to access the complete crossover other
than for
two dimensional films \citelabel{\FerdFisher}[\cite{FerdFisher}].

The format of the paper is as follows: in section 2  we analyze the connection
between the specific heat and the vertex functions of a
Landau-Ginzburg-Wilson effective Hamiltonian. By choosing as mass parameter
$t_B=\L^2{(T-T_c)\over T}$
we include all the non-analytic dependence of the specific heat in
the vertex function $\G^{(0,2)}_B$.
We then present a renormalization of the theory as a reparametrization
through normalization conditions on certain renormalized vertex functions
using a fiducial correlation length as our RG scale.
Section 3 is devoted to perturbative calculations.
In particular we calculate the specific heat and a specific heat effective
exponent to one loop.
In section 4, by matching to the known asymptotic exponents for a three
dimensional Ising film, we access the crossover between power law behaviour at
the three dimensional end and logarithmic behaviour at the two dimensional end.
The results in sections 3 and 4 are illustrated in the figures.
The paper ends with our conclusions.

\beginsection{Renormalization of the Specific Heat}
We consider an $O(N)$ symmetric order parameter described by the
``microscopic'' Landau-Ginzburg-Wilson Hamiltonian
\eqlabel{\ham}
$$H[\vf_{B}]=\int_0^L\int
d^{d}x\left({1\o2}(\nabla\vf_{B})^2+{1\o2}m^2_{B}\vf_{B}^2+
{1\o2}t_{B}(x)\vf_{B}^2
+{\l_{B}\o4!}\vf_{B}^{4}-H_B(x)\vf_B\right)\no$$
which describes a $d$ dimensional film geometry of thickness $L$.
The variable $t_{B}$ when taken to be homogeneous
has analytic dependence on temperature $T$,
and we choose its origin
to be the critical temperature of the film.  Hence $m_{B}^2$ is determined by
the difference between the $L$ dependent critical temperature and the mean
field critical temperature, i.e. the temperature at which the potential
in (\docref{ham}) acquires a non-zero minimum. $\l_{B}$ is assumed to
be temperature independent.
The subscript $B$ refers to bare parameters as distinct from
renormalized parameters which will be introduced below.
We will restrict attention to the case when the
film also exhibits a phase transition and consider
$3\leq d\leq4$ for $N=1$, and $3<d\leq4$ for $N>1$.
We will present results for periodic, antiperiodic and Dirichlet boundary
conditions.
Note that in the case of periodic boundary conditions our
results for $N=1$ could equally well be reinterpreted to describe the quantum
to classical crossover
of an Ising model in a transverse magnetic field $\G$, where now
$t_B=\G-\G_c(T)$ and $L={\hbar/T}$
\citelabel{\Hertz}[\cite{Hertz}]. However,
in this paper we restrict our considerations to the film geometry.

The partition function
for the the model (\docref{ham}) is given by the path integral
\eqlabel{\partitionfn}
$$Z=\int[d\vf_{B}]e^{-H[\vf_B]-{V\over T}F^b} . \no$$
The free energy density is $F=-{T\over V}\ln Z=F^b-{T\over V}\ln Z_{LGW}$,
where
$V$ is the volume and  $F^b$ is the background free
energy density obtained after coarse graining from the underlying microscopic
degrees of freedom to those of the effective field theory description in terms
of the
LGW Hamiltonian (\docref{ham}), $Z_{LGW}$ being the partition function of
this Hamiltonian. $F^b$ is assumed to be an analytic function of
the thermodynamic variables.
The internal energy density is
$$U=F-T{\p F\over \p T}$$
and the specific heat, by definition $\p U/\p T$, is given by
\eqlabel{\specificheat}
$$C=-T^2{\p^2F\o\p T^2} . \no$$
The assumption in working with this LGW Hamiltonian is that
the only one of its parameters to retain a dependence on
temperature is the mass parameter $t_B$.
Thus the internal energy density
$$U=U^b-{T^2\over2V}\int d^dx{\p t_B(x)\over \p T}G^{(0,1)}\no$$
and the specific heat
$$C=C^b-{1\over2V}\int d^dx({\p\over \p T}T^2{\p t_B(x)\over \p T})G^{(0,1)}(x)
+{T^2\over4V}\int d^dx \int d^dy{\p t_B(x)\over \p T}G^{(0,2)}(x,y){\p
t_B(y)\over \p T}\no$$
where
$$G^{\ss (0,1)}(x)=\langle\varphi^2(x)\rangle\quad\hbox{ and }\quad
G^{\ss
(0,2)}(x,y)=\langle\varphi^2(x)\varphi^2(y)\rangle-\langle\varphi^2(x)\rangle\langle\varphi^2(y)\rangle . \no$$

Concentrating on $T>T_c$, where $<\varphi>=0$, and denoting
$$\G[t_B]=-\ln Z_{LGW}\no$$
we have that
$$\G^{(0,1)}(x)={1\over2}G^{(0,1)}(x)\qquad\hbox{ and }
\qquad \G^{(0,2)}(x,y)=-{1\over4}G^{(0,2)}(x,y) . \no$$
So for homogeneous $t_B$  we find
$$U=U^b-T^2{\p t_B\over \p T}\G_B^{(0,1)}\no$$
$$C=C^b-({\p \over \p T}T^2{\p t_B\over \p T})\G_B^{(0,1)}
-T^2{({\p t_B\over \p T})}^2\G_B^{(0,2)},\no$$
where $\G_B^{(0,1)}$ and $\G_B^{(0,2)}$ are to be evaluated at zero
external momentum.

If we wish to incorporate all of the non-analytic dependence of the
internal energy and the specific heat into $\G^{(0,1)}$ and $\G^{(0,2)}$
respectively, then a natural choice of  the dependence of $t_B$
is
\eqlabel{\tboftemp}
$$t_B=\Lambda^2\ {(T-T_c)\over T}\no$$
where $\Lambda$ is a microscopic mass scale. In the vicinity of the
critical temperature the results with this  variable will be the
same as those obtained with the linear measure $\Lambda^2\ {(T-T_c)\over T_c}$.
With the choice (\docref{tboftemp}) the  internal energy density becomes
$$U=U^b-\Lambda^2T_c\G_B^{(0,1)}\no$$
and the specific heat is given by
$$C=C^b-{\Lambda^2\over T^2}\G_B^{\ss (0,2)}.\no$$
For an $O(N)$ model $G_B^{(0,2)}$ is manifestly positive and either
diverges or goes to zero at the critical temperature according to the value of
$N$.
Thus we anticipate
that $\G_B^{(0,2)}$ should diverge to $-\infty$ or vanish at the critical
temperature.
Our problem is therefore to calculate $\G_B^{(0,2)}$.

The correlation length in the transverse dimensions, $\xi_L=m^{-1}$,
we define via the second moment of the two point function, $G^{(2)}$.
On Fourier transforming $\xi_L$ is  obtained from
\eqlabel{\physicalmass}
$$m^{2}=\left.{\G_B^{(2)}(p,t_B(m),\l_B,L)
\over \p_{p^2}\G_B^{(2)}(p,t_B(m),\l_B,L)}\right\vert_{p^2=0}
\no$$
where $p$ is the transverse momentum and $t_B(m)$ is that bare
mass parameter which produces the inverse correlation length $m$.
The origin for the variable $t_B(m)$ is specified by requiring that
\eqlabel{\tbareorigin}
$$\G_B^{(2)}(0,0,\l_B,L)=0
\no$$
which insures that $t_B$ is proportional to
$T-T_c(L)$ as the critical temperature is approached.
Changing the mass parameter $t_B$, by changing the
temperature in (\docref{tboftemp}), allows us to tune the correlation
length. Note that the physical correlation length of the
film geometry (\docref{physicalmass}) depends on $L$ and will be infinite
at the film critical temperature $T_c(L)$.

We will define renormalized parameters by
\eqlabel{\trenorm}
$$t(m,\k)=Z_{\varphi^2}^{-1}(\k)t_B(m)\qquad \hbox{ and }\qquad
\l(\k)=Z_\l(\k)\l_B \no$$
and renormalized vertex functions by
\eqlabel{\vertexren}
$$\G^{(N,M)}(m,\k)=Z_{\varphi}^{{N\over2}}(\k)Z_{\varphi^2}^{M}(\k)\G_B^{(N,M)}(m)
+\delta_{N0}\delta_{Mn}A^{(n)}(\k) \qquad n=0,1,2\no$$
which is just a reparametrization of the original theory, where $\k$ is an
arbitrary renormalization scale.

Contrary to the renormalization of
other vertex functions, like $\G_B^{(2)}$ and $\G_B^{(4)}$,
the vertex functions $\G_B^{(0,n)}$ ($n=0,1,2$) have to be renormalized
additively via
\eqlabel{\adren}
$$\G^{(0,n)}=Z^n_{\vft}\G^{(0,n)}_B+A^{(n)}\no$$
$\G$ and $\G^{(0,1)}$ determine the Gibbs free
energy  density and the energy density  of the LGW Hamiltonian respectively
\footnote{*}{For homogeneous $t_B$, $H_B=0$ and $T>T_c$  we use the convention
$\G={1\over V}\ln Z_{LGW}$} .

Before discussing the renormalization of $\G^{(0,n)}$ for $n=0,1,2$
we will specify the $Z$'s associated with the reparametrization
(\docref{trenorm}) and (\docref{vertexren}). Here we will restrict ourselves to
$T>T_c(L)$.
The case of $T<T_c(L)$  will be considered in conjunction with
crossover amplitude ratios elsewhere.
For $T>T_c(L)$ the conditions which
specify our $Z$'s are
\eqlabel{\Zphi}
$$Z_{\varphi}^{-1}=\left.\p_{p^2}\G_B^{(2)}(p,t_B(\k),\l_B,L)\right\vert_{p^2
=0}\no$$
\eqlabel{\Zphis}
$$Z_{\varphi^2}^{-1}=\left. {\G_B^{(2,1)}(p, t_B(\k), \l_B, L)\over
\p_{p^2}\G_B^{(2)}(p,t_B(\k),\l_B,L)}\right\vert_{p^2 =0} \no$$
\eqlabel{\Zlam}
$$ Z_{\l}={\G_B^{(4)}(0, t_B(\k), \l_B, L)\over \l_B}\no$$
where the relation between $t_B(\k)$ and $\k$ is specified by
\eqlabel{\refmass}
$$\k^2=\left. {\G_B^{(2)}(p,t_B(\k),\l_B,L)
\over \p_{p^2}\G_B^{(2)}(p,t_B(\k),\l_B,L)
} \right\vert_{p^2=0}
\no$$
and the origin of $t_B$ is fixed by (\docref{tbareorigin}).
Note that the $Z$'s are obtained from the vertex functions of the
system specified at an arbitrary, fiducial, transverse correlation length
$\k^{-1}$,
as opposed to the  correlation length of interest, $m^{-1}$.
Furthermore the conditions are all $L$ dependent. As has been emphasized on
previous occasions [\cite{NucJphysa},\cite{EnvfRG}]
such ``environmentally friendly''  conditions are
essential in order to obtain a perturbatively controllable description
of the finite size crossover.

We define the Wilson functions as the logarithmic derivatives
\eqlabel{\wilsongf}
$$\gf={1\over Z_{\varphi}}\k{d Z_{\varphi}\over d\k}\no$$
\eqlabel{\wilsongft}
$$\gft=-{1\over Z_{\varphi^2}}\k{d Z_{\varphi^2}\over d\k}\no$$
\eqlabel{\wilsongl}
$$\gl={1\over Z_{\l}}\k{d Z_{\l}\over d\k} . \no$$
The Wilson functions $\gft$, $\gf$ and $\gl$ are explicitly $L$ dependent
and interpolate between those of a $d$ and $d-1$ dimensional $O(N)$
model in the limits
$\k L\ra\i,\ \k\ra0$ and $\k L\ra0,\ \k\ra0$ respectively.

The invariance of the bare vertex functions,
$\Gamma^{(N,L)}_B$, under the one parameter group of
reparametrizations indexed by the
arbitrary renormalization scale $\k$
(they don't know which reference
correlation length $\k^{-1}$ will be picked to define the reparametrization)
yields the RG equation
\eqlabel{\rgee}
$$\k{d\Gnl\o d\k}+(L\gft-{N\o2}\gf)\Gnl=
\delta_{\s N0}\delta_{\s Ln}B^{(n)}\no$$
where $n=0,1,2$.
The equation is inhomogeneous for the three vertex
functions $\G$, $\G^{(0,1)}$ and $\G^{(0,2)}$, where the ``source'' term
\eqlabel{\sou}
$$B^{(n)}=\k{dA^{(n)}\o d\k}+n \gft A^{(n)}\no$$
is finite order by order in the loop expansion.

The relationship between temperature and $\k$ can be obtained
by using
\eqlabel{\integ}
$$\G^{(2)}(t)=\int_0^t\G^{(2,1)}(t')dt' \no$$
and conditions (\docref{Zphi}-\docref{refmass}) with the definitions
of the Wilson functions (\docref{wilsongf}-\docref{wilsongl}) to find
\eqlabel{\eqofsta}
$$t(m,\k)=\kappa^2\int_0^{m}{dx\o x}(2-\gf)e^{\int_\k^x(2-\gft){dy\over y}} .
\no$$
We see that
$$\k{dt(m,\k)\over d\k}=\gft t(m,\k) .
\no$$
An important feature of the above is that  the determination of
$\G^{(2)}$ by integrating $\G^{(2,1)}$ allows us to
bypass the need to determine $m_B^2$ perturbatively.

In terms of the renormalized vertex functions, the conditions
(\docref{Zphi}-\docref{refmass}) are equivalent to
\eqlabel{\nctwo}
$$\left.\p_{p^2}\G^{(2)}(p,t(\k,\k),\l,L,\k)\right\vert_{p^2=0}=1\no$$
\eqlabel{\ncthree}
$$\G^{(2,1)}(0,t(\k,\k),\l,L,\k)=1\no$$
\eqlabel{\ncfour}
$$\G^{(4)}(0,t(\k,\k),\l,L,\k)=\l .\no$$
\eqlabel{\ncone}
$$\G^{(2)}(0,t(\k,\k),\l,L,\k)=\k^2 . \no$$
We could have replaced (\docref{ncthree}) by the condition
\eqlabel{\ncfive}
$$t(\k,\k)=\k^2 .
\no$$
This condition  together with (\docref{ncone}) determines a
multiplicative renormalization of $t_B$, and of $\vft$
insertions via a renormalization
function $Z_t$. The two renormalization
functions $Z_t$ and $Z_{\vft}$ are
different, the latter being determined
by (\docref{ncthree}).
The quantity  $\gamma_t=-{d\ln Z_t\o d\ln\k}$ is an analog of $\gft$,
however, the problem with
implementing a condition such as (\docref{ncfive}) in perturbation
theory is that the resulting $Z_t$ involves
diagrams with massless propagators, some of which are strictly infinite
even after the introduction of an ultraviolet cutoff.

Defining an effective exponent $\nu\ef={d\ln\xi_L^{\s -1}\o d\ln t}$, one finds
\eqlabel{\neff}
$$\nu\ef=(2-\gamma_t)^{-1}={\int_0^{\ss{\xi_L^{-1}}}{dx\o x}(2-\gf)
e^{\int_{\xi_L^{-1}}^x{dy\o y}(2-\gft)}\o(2-\gf)} . \no$$
One can also define what we term a floating exponent, $\nu_{\ss
f}=(2-\gft)^{-1}$.
As near a fixed point $\gf$ and $\gft$ go to constants we can see from
(\docref{neff}) that $\gamma_t\ra\gft$, hence  both the true effective
exponent and the floating exponent interpolate between the same two
asymptotic values. This may not be evident perturbatively.  One can think of
the floating  exponents evaluated in environmentally friendly RG improved
perturbation theory as
approximations to the true effective exponents [\cite{EnvfRG}].
Another way of thinking about them is from the point of view
of a redefined temperature variable in the following way: if one defines
$t'=tf(t)$ and
$\nu_{\ss f}={d\ln\xi_L^{-1}\o d\ln t'}=(2-\gft)^{-1}$ one finds that
\eqlabel{\coord}
$${d\ln f\o d\ln t}+1={(2-\gft)\o(2-\gf)}
\int_0^{\ss{\xi_L^{-1}}}{dx\o x}(2-\gf)e^{\int_{\xi_L^{-1}}^x{dy\o y}(2-\gft)}
. \no$$
Near a fixed point $f\ra1$ hence $t'\ra t$.

The solution of the RG equation (\docref{rgee}) for the specific heat is
\eqlabel{\scasol}
$$\eqalign{\Gamma^{(0,2)}(t(m,\k),\l(\k), L, \k)=&e^{2\inr {dx\o x} \gft}
\Gamma^{(0,2)}(t(m,\r\k), \l(\r\k), L,\k\r)\cr
&\qquad - \inr
{dx\o x} B^{(2)}(x) e^{2\int_{1}^x {dy\o
y} \gft} . } \no$$
Reparametrization
invariance is now manifest in the fact that the left hand side of
(\docref{scasol}) is independent of $\r$, the latter being just an arbitrary
rescaling
of $\k$.

We will now discuss some  possible
normalization conditions for $\Got$, thus specifying $A^{(2)}(\k)$.
One possible choice is
the normalization condition
\eqlabel{\vannc}
$$\Got(t(\k,\k),\lambda(\k),L,\k)=0\no$$
which is equivalent to
$$A^{(2)}(\k)=-Z_{\vft}^{2}\Got_B(t_B(\k),\l_B,L) .
\no$$
The advantage of this condition is that all the ``physics'', in the sense of
the effects of all fluctuations, is now purely in the inhomogeneous
term. The normalization condition (\docref{vannc}), however, is natural as the
$\G^{(0,2)}$ does indeed vanish in the mean field regime, or at least goes to a
constant which can be chosen to be zero. Neglecting the inhomogeneous term
$\G^{(0,2)}$ being zero is
then an invariant statement with respect to RG transformations.

A methodology which avoids some of the pitfalls of additive renormalization is
to relate $\G^{(0,2)}$ to the correlation function $\G^{(0,3)}$, the advantage
of this approach being that the latter is multiplicatively
renormalizable in $d<6$.
We have the analog of  (\docref{integ})
\eqlabel{\integsp}
$$\G^{(0,2)}(t)-\G^{(0,2)}(t_i)=\int_{t_i}^t\G^{(0,3)}(t')dt' . \no$$
Using the relation
\eqlabel{\rggzth}
$$\G^{(0,3)}(t(m,\k),\l(\k),L,\k)=e^{3\int^{\r}_1{dx\o
x}\gft}\G^{(0,3)}(t(m,\r\k),\l(\r\k),L,\k\r)\no$$
and the relation between the correlation length and the temperature
(\docref{eqofsta}) one finds
\eqlabel{\gamzerth}
$$\G^{(0,2)}(t(m,\k))=\k^{d-4}\int_{\i}^m{dx\o x}(2-\gf)
e^{\int^x_1(2\gft-4+d){dy\o y}}\bar\G^{(0,3)}(x)\no$$
where we have normalized $\G^{(0,2)}$ to vanish in the mean field limit
and
$$\bar\G^{(0,3)}(m)={\G^{(0,3)}{\G^{(2)}}^3\o{\G^{(2,1)}}^3m^d} .
\no$$
It is not difficult to show that in fact (\docref{gamzerth}) is exactly the
same as
the expression (\docref{scasol}) obtained from the additive renormalization
prescription
with the normalization condition (\docref{vannc}) at $t(\infty,\infty)$.

\beginsection{Perturbative calculations}
We begin this section by analysing the $\beta$ function for the coupling, as
we will perturbatively expand all other functions in terms of the solution of
this equation .
In terms of the floating coupling $h$ [\cite{NucJphysa}], chosen to be the
leading term in the perturbative series for $\gl$, one finds, for $\r {dh\over
d\r}=\beta(h,z)$,
to one loop
\eqlabel{\betahz}
$$\beta(h,z)=-\e(z) h+h^2.\no$$
The function $\e$, in an obvious diagrammatic notation, is
$$\e(z)={6\k^4\bfour\over\itro}-2  \  ,\no$$
depends on $d$ and
$z={\r\k L}$ but is independent of $N$.
We take the solution of (\docref{betahz})
\eqlabel{\betahNol}
$$h(z)={e^{-\int_{z_{\ss 0}}^{z}\e(x){dx\o x}}\o{h_0^{-1}
-\int_{z_0}^{z}e^{-\int^x_{z_{\ss 0}}\e(y){dy\o y}}{dx\o x}}}\no$$
as our perturbation parameter.

After solving the equation we specify the arbitrary scale $\r$
to be $\r={1\over\k\xi_L}$ and relate it to temperature via
(\docref{eqofsta}) whereupon $z$ becomes $L/\xi_L$. In (\docref{betahNol})
the initial coupling is then taken to be at a ``microscopic'' scale $\k$. For
$d<4$ this microscopic scale can be sent to infinity and a universal
floating coupling, the separatrix solution  $h(z)={4 z^2\itro\o\bub}$
obtained [\cite{EnvfRG}]. Of course, if one is interested in
corrections to scaling, as is usually the case in comparing with
experimental data, then $\k$ should be left finite and fitted to the data.

For periodic boundary conditions one finds
$$\e(z)=5-d-(7-d){{\displaystyle\sum_{n=-\i}^{\i}{4\pi^2n^2\over z^2}
\left(1+{4\pi^2n^2\over z^2}\right)^{d-9\over2}}
\over{\displaystyle\sum_{n=-\i}^{\i}\left(1+{4\pi^2n^2\over
z^2}\right)^{d-7\over2}}}\no$$
and the separatrix coupling
\eqlabel{\perh}
$$h(z)=(5-d){{\displaystyle\sum_{n=-\infty}^{\infty}{(1+{4\pi^2 n^2\o
z^2})}^{(d-7)\o2}}
\o{\displaystyle\sum_{n=-\infty}^{\infty}{(1+{4\pi^2 n^2\o
z^2})}^{(d-5)\o2}}} .\no$$
For $d=3$ the results are particularly simple
\eqlabel{\htd} $$h(z)=1+{z\o\sinh z}\no$$
\eqlabel{\epstd}
$$\e(z)=1+{z^2\coth({z\o2})\o\sinh z+z}\no$$
where, of course, we are now restricted to $N=1$.

We present here the corresponding results for Dirichlet and antiperiodic
boundary conditions. For Dirichlet boundary conditions
$$\e(z)=5-d-(7-d){{\displaystyle\sum_{n=1}^{\i}{\pi^2(n^2-1)\over z^2}
\left(1+{\pi^2(n^2-1)\over z^2}\right)^{d-9\over2}}
\over{\displaystyle\sum_{n=1}^{\i}\left(1+{\pi^2(n^2-1)\over
z^2}\right)^{d-7\over2}}}\no$$
and for the separatrix coupling
$$h(z)=(5-d){{\displaystyle\sum_{n=1}^{\infty}}{(1+{\pi^2(n^2-1)\o
z^2})}^{(d-7)\o2}
\o{\displaystyle\sum_{n=1}^{\infty}}{(1+{\pi^2(n^2-1)\o
z^2})}^{(d-5)\o2}} . \no$$
For antiperiodic boundary conditions one finds
$$\e(z)=5-d-(7-d){{\displaystyle\sum_{n=-\i}^{\i}{\pi^2n(n+1)\over
z^2} \left(1+{\pi^2n(n+1)\over z^2}\right)^{d-9\over2}}
\over{\displaystyle\sum_{n=-\i}^{\i}\left(1+{\pi^2n(n+1)\over
z^2}\right)^{d-7\over2}}}\no$$
and finally the separatrix coupling
$$h(z)=(5-d){{\displaystyle\sum_{n=-\i}^{\infty}}{(1+{\pi^2n(n+1)\o
z^2})}^{(d-7)\o2}
\o{\displaystyle\sum_{n=-\i}^{\infty}}{(1+{\pi^2n(n+1)\o
z^2})}^{(d-5)\o2}} . \no$$

For $d=3$ the results are once again very simple. For the Dirichlet case
\eqlabel{\dir}
$$\e(y)=1+{3\pi^2\o
y^2}+2(1+{\pi^2\o y^2}) {{({y^2\o\sinh^2y}-{\tanh y\o
y})}\o{(1+{2y\o\sinh 2y}-2{\tanh y\o y})}}\no$$
where $y=(z^2-\pi^2)^{1\o2}$.
Even though $y(z)$ has a branch point $h(z)$ is analytic in $z$.
The separatrix coupling is
\eqlabel{\dirh}
$$h(y)=(1+{\pi^2\o y^2}){(1+{2y\o \sinh 2y}-{2\tanh y\o y})\o (1-{\tanh
y\o y})} . \no$$
The corresponding results for antiperiodic boundary conditions are
\eqlabel{\anti}
$$\e(y)=1+{3\pi^2\o y^2 }-{(y^2+\pi^2)\tanh(y/2)\o(\sinh y-y)}\no$$
and
\eqlabel{\antih}
$$h(y)=(1 + {\pi^2\o y^2})(1 - {y\o\sinh y}) . \no$$

The Wilson function $\gft$ is given by
\eqlabel{\gfthz}
$$\gft(h,z)={(N+2)\o (N+8)}h\no$$
whilst $\gf=0$ to one loop. Two loop Pad\'e resummed expressions for the
Wilson functions and the floating coupling, for the case of periodic boundary
conditions,
can be found in [\cite{EnvfRG},\cite{PRL}].
Substituting any of the above floating couplings into
(\docref{gfthz}) yields $\gft$ for the three different types of boundary
condition.
As mentioned corrections to scaling can easily be included. For example for
$d=3$ and periodic boundary conditions
\eqlabel{\nonuni}
$$h^{-1}(z)={z\sinh({z\over 2})^2\over\sinh z+z}({1\over h(z_{0})}{\sinh
z_0+z_0\over z_0\sinh({z_{0}\over2})^2} -2{\coth({z_{0}\over2})\over
z_{0}})+{\sinh z\over \sinh z+z} . \no$$

Turning now to $\Got$, up to two loop order and once again in an
obvious diagrammatic notation (note that we have made the diagrams
dimensionless by pulling out an overall scale)
 $\G^{(0,2)}_B$ is given by
\eqlabel{\diash}
$$\G^{(0,2)}_B = -{N\o2}(\r\k)^{d-4} \Bigl[\bub-\l^{}_B
\k^{d-4}{(N+2)\o6}\bub^2\Bigl] \no$$
the two loop graph with ``tadpole''  having been absorbed into the one loop
propagator
by the replacement of $t_B$ with $\r\k$ using (\docref{refmass}).  Implementing
the normalization condition
(\docref{vannc}) one finds that
\eqlabel{\source}
$$ B^{(2)}=-2N\k^2\itro|_{n.p}  \no$$
where the subscript denotes that the diagram is evaluated at the normalization
point. Thus we see that the one and two loop expressions for $B^{(2)}$ in terms
of renormalized quantities are identical. Explicitly to one loop for periodic
boundary conditions one finds
$$B^{(2)}=-{N\o L}{\G({7-d\o2})(\r\k)^{d-5}\o (2\pi)^{{d-1\o2}}}
\sum_{n=-\i}^{\i}(1+{4\pi^2n^2\o L^2\k^2\r^2})^{d-7\o2} . \no$$
$\G^{(0,2)}$ is thus found by substituting (\docref{source}) and
(\docref{gfthz}) into (\docref{scasol}) to obtain
\eqlabel{\spht}
$$\eqalign{\Got(t,\l,L,\k)=&{N\o 2L\k}{\G({7-d\o2})\k^{d-4}\o
(2\pi)^{{d-1\o2}}}\cr
&\int^{\r}_1{dx\o x}x^{d-5}
\sum_{n=-\i}^{\i}(1+{4\pi^2n^2\o L^2\k^2x^2})^{d-7\o2}
e^{2\int^x_1{\left({N+2\o N+8}\right)h{dy\over y}}}\cr}\no$$
where the arbitrary scale $\r$, as before, is associated directly with the
inverse correlation length. Thus we calculate
the specific heat and other physical quantities directly in terms of the finite
size correlation length. Equation (\docref{eqofsta}) relating $\xi_L$ to $L$
and
$t$ provides a parametric representation of physical quantities in terms
of $t$.

In the limit
$\r\ra0$ only the $n=0$ term in the sum is important and one finds
\eqlabel{\aslim1}
$$\Got\ra-{N(N+8)\o 2(4-N)L\k}{\G({5-d\o2})\k^{d-4}\o (2\pi)^{{d-1\o2}}}
\r^{d-5+2(5-d){\left({N+2\o N+8}\right)}} . \no$$
In the same limit one finds $\r\ra{({t\o\k^2})}^{\nu_{\ss d-1}}$,
$\nu_{\ss d-1}=(2-{\left({N+2\o N+8}\right)}(5-d))^{-1}$ being the $d-1$
dimensional correlation length exponent. Hence
\eqlabel{\aslim2}
$$\Got\ra-{N(N+8)\o 2(4-N)L\k}{\G({5-d\o2})\k^{d-4}\o (2\pi)^{{d-1\o2}}}
({t\o\k^2})^{-\alpha_{\ss d-1}}\no$$
where $\alpha_{\ss d-1}={5-d-2(5-d){\left({N+2\o N+8}\right)}
\o{2-{\left({N+2\o N+8}\right)}(5-d)}}$ is the $d-1$ dimensional specific heat
exponent. Similarly, in the limit $L\k\r\ra\i$, $\r\ra0$ the sum can be
converted to an integral and one finds that
\eqlabel{\aslim3}
$$\Got\ra-{N(N+8)\o 2(4-N)}{\G({4-d\o2})\k^{d-4}\o (2\pi)^{d/2}}
\r^{d-4+2(4-d){\left({N+2\o N+8}\right)}}\no$$
and $\r\ra{({t\o\k^2})}^{\nu_d}$ where $\nu_d$ is the $d$ dimensional
correlation length exponent. In the bulk limit
\eqlabel{\aslim4}
$$\Got\ra-{N(N+8)\o 2(4-N)}{\G({4-d\o2})\k^{d-4}\o (2\pi)^{d/2}}
({t\o\k^2})^{-\alpha_{d}}\no$$
where $\alpha_{d}={4-d-2(4-d){\left({N+2\o N+8}\right)}
\o{2-{\left({N+2\o N+8}\right)}(4-d)}}$ is the $d$ dimensional specific heat
exponent. Thus we see that the specific heat crosses over precisely between
the expected $d$ and $d-1$ dimensional asymptotic forms.

Note that the amplitude of $\G^{(0,2)}$ in the above expressions
appears to diverge at $N=4$ this is an artifact of the one loop approximation.
What actually happens is that for  $d$ between two and four there is
some value of $N$ for which $\alpha(N,d)=0$, at this value of $N$ and $d$
we expect the specific heat to have a logarithmic dependence on $t$. For $N=1$
this
occurs at $d=2$, however, at one loop the value appears to be independent
of $d$ and occurs at $N=4$, which is the relevant value for $d=4$.

A plot of the
specific heat as a function of correlation length is shown in Fig. 1
for a three dimensional Ising film with periodic boundary conditions.
The effective specific heat exponent defined as
$$\alpha\ef=-{d\ln C\over d\ln t}\no$$
is plotted in Fig. 2 for the same model.  Note that in this approximation
the asymptotic two dimensional
value of $\alpha\ef$ is $0.5$ as opposed to the exact value of zero, obtained
from the solution of the two dimensional Ising model. This is a weakness of the
perturbative
approach which effects the specific heat exponent in a particularly acute
manner. In the next section by matching to the known asymptotic exponents of
the model we investigate
the more realistic behaviour.
%\midinsert
%\epsffile{specg1.ps}
%\endinsert
In the case of
a four dimensional $O(N)$ film, in the limit ${L\o\xi_L}\ra\i$,
$\xi_L\ra\i$, one finds that
\eqlabel{\aslim}
$$\Got\ra-{N\o16\pi^2}\left({N+8\o4-N}\right)
\left\vert{1\o2}\ln{t\o\k^2}\right\vert^{4-N\o N+8}\no$$
in accordance with known results. Fig.'s 3 and 4 show plots of
the specific heat and $\alpha\ef$ for the four dimensional Ising film.
Note the presence in the figures of  logarithmic tails at the four
dimensional end as described by (\docref{aslim}).
Fig. 5 shows a comparison
of $\alpha\ef$ for a three dimensional Ising film with Dirichlet and
antiperiodic boundary conditions.
%\midinsert
%\epsffile{specg2.ps}
%\endinsert
Additionally the result for the Gaussian model is plotted  in Fig. 6
with periodic boundary conditions.
%\midinsert
%\epsffile{specg3.ps}
%\epsffile{specg4.ps}
%\endinsert

\beginsection{Crossover to Logarithmic Behaviour in a Three Dimensional Ising
Film.}
In this section we will consider the crossover between three and two dimensions
for an Ising model in a way that is capable of accessing the logarithmic
behaviour characteristic of the two dimensional specific heat.
For the two dimensional
Ising model $\alpha=2-\nu d=0$. The consequent logarithmic behaviour
of the specific heat is thus due to a competition between $\nu$ and $d$.
For $d=2$
the correlation length exponent $\nu=1$, hence $\alpha=0$.
Now for a three
dimensional  Ising film with periodic boundary conditions,
at one loop the crossover is governed
by the floating coupling $h=1+{z\over \sinh z}$.
This implies a crossover for $\nu_{eff}$ between $1/6$ and $1/3$.
By far the biggest error involved in evaluating
crossover functions is associated with the values of the asymptotic exponents
themselves.
With this in mind one is inclined to try to match the scaling function to the
asymptotic exponents.
This can very simply be done in the case at hand by writing
$h=A+{Bz\over \sinh z}$ where now the constants $A$ and $B$ will
be determined by
demanding that as $z\ra0$, $\nu_{eff}\ra1$ and that as $L\ra\i$, $z\ra0$ one
finds
$\nu_{eff}\ra0.630$. The values $1$ and $0.630$ are the exact two dimensional
and three dimensional 6-loop Borel resummed
\citelabel{\BkNiM} [\cite{BkNiM}] exponents
respectively. Thus one finds
that $A= 1.238$ and $B=1.762$.

In Fig. 7 we plot $\alpha_{\ef}$ as a function of $\ln z$ by substituting our
ansatz
for $h$ into (\docref{spht}).
Note the logarithmic tail as the two dimensional critical region is approached.
More interestingly, there is a pronounced bump in the curve which is absent
in the one-loop approximation.
This arises due to a competition between the effects of $\nu_{\ef}$ and the
effective dimensionality $d_{\ef}$[\cite{EnvfRG}].
The bump remains even if one uses a completely different interpolating
function such as $h=A+{Bz\o1+z}$, though its amplitude and width vary
somewhat.
%\midinsert
%\epsffile{specg5.ps}
%\endinsert
In Fig. 8 we plot analogous results for the case
of Dirichlet and antiperiodic boundary conditions. Once again
the bump is clearly present. In the case of Dirichlet conditions
however there is also a dip before the bump is reached. Based on
previous experience of the behaviour of effective exponents with
Dirichlet boundary conditions [\cite{EnvfRG}]
this is not totally unexpected.
In Fig.'s 9 and 10 we have  used instead of the universal floating
coupling the coupling
(\docref{nonuni}). There is now a double crossover; firstly between
mean field theory and
the three dimensional asymptotic exponent and then to the asymptotic
behaviour of the two dimensional exponent. In Fig. 9 we plot the result
for the case where
we do not match to the exact two dimensional exponent and in
Fig. 10 the result with
matching. The asymptotic three dimensional regime would most
probably be much narrower
than that shown. This can be very easily modeled by
adjusting the initial condition for the
RG flow. In the case at hand, we have, for the sake of clarity, and to
emphasize
the double crossover, left it large. It is clear from the figure how the
effective
exponent
would be modified as the well developed three dimensional universal
regime is narrowed.

\beginsection{Conclusions}
In this paper, using environmentally friendly renormalization,
we have treated the finite size crossover of the
specific heat of an $O(N)$ model in a $d$-dimensional
film geometry. For $N>1$ we
considered $3<d\leq4$, and for $N=1$, $3\leq d\leq4$. We derived
expressions for the specific heat and an effective critical
exponent $\alpha\ef$ that were completely regular across the entire
crossover, the expansion parameter for the perturbative
series being the floating coupling $h$. We considered periodic, Dirichlet and
antiperiodic boundary conditions.

For the crossover from three to two dimensions of an Ising film
we saw that one loop answers in the asymptotic two dimensional
regime were quite poor. As is known, generally speaking,
perturbation theory becomes more
unreliable as one goes to lower dimensions. For the specific heat
the problem is particularly acute as the one loop effective specific
heat exponent was seen to be monotonically increasing
whereas, as we know from the
solution of the exact two-dimensional Ising model, the two dimensional
specific heat exponent is strictly less than the three dimensional one.
Hence we could say that the
one loop approximation is failing to capture a qualitative feature
of the crossover in this case. To circumvent this problem, and in
the knowledge that the dominant source of error in calculating
scaling functions is the uncertainty in the asymptotic critical
exponents, we took a more pragmatic line by making an ansatz for
the floating coupling constant so as to be able to asymptotically match the
``known'' two and three dimensional correlation length exponents.
By so doing we were able to access in a very simple way the
crossover between power law and logarithmic behaviour in the
asymptotic regime, finding that the resultant crossover curve
had a very interesting bump. We also analyzed the crossover to mean field
theory thereby accessing a double crossover governed by three
different fixed points. Our global, environmentally
friendly RG captured all of these fixed points in one uniform
approximation scheme.

The crossover between
three and two dimensions for $N>1$, and in particular for $N=2$, are potential
problems that could be analyzed using the techniques of this paper. The
latter being a problem of longstanding interest for
experiments with liquid helium confined to a film geometry \citelabel{\Expts}
[\cite{Gasparini},\cite{Expts}]. We hope to return to these issues in the
future.

\vfill\eject
\line{\bf \big Acknowledgements.\hfill}
We would like to thank Prof. Michael Fisher for his interest and several
very useful discussions especially relating to the matching of asymptotic
exponents. CRS was supported by an EU Human Capital and Mobility
Fellowship.
\bigskip
\line{\bf \big References. \hfill}

\item{[\cite{NucJphysa}]} D.\ O'Connor and C.R.\ Stephens,
  {\it Nucl.\ Phys.}\ {\bf B360} (1991) 297;
  {\it J. Phys.}\ {\bf A25} (1992) 101.

\item{[\cite{EnvfRG}]} D.\ O'Connor and C.R.\ Stephens,
  {\it Int.\ J.\ Mod.\ Phys.}\ {\bf A9} (1994) 2805.

\item{[\cite{Procroys}]} D.\ O'Connor and C.R.\ Stephens,
 {\it Proc.\ Roy.\ Soc.}\ {\bf  444A}, (1994), 287.

\item{[\cite{EqnofState}]} F.\ Freire, D.\ O'Connor and C.R.\ Stephens,
 {\it J.\ Stat.\ Phys.}\ {\bf 74}, (1994) 219.

\item{[\cite{PRL}]} D.\ O'Connor and C.R.\ Stephens,
 {\it Phys. Rev. Lett.}\ {\bf 72}, (1994) 506.

\item{[\cite{Gasparini}]}T.P.\ Chen and F.M. Gasparini,
{\it Phys.\ Rev.\ Lett.}\ {\bf 40}, (1978) 331;\hfill\break
F.M.\ Gasparini, G.\ Agnolet and J.D. Reppy,
{\it Phys.\ Rev. }\ {\bf B 29}, (1984) 128.

\item{[\cite{Binder}]}K.\ Binder, {\it Z.\ Phys.}\ {\bf B 43}, (1981) 119.

\item{[\cite{RudGuoJas}]}J. Rudnick, H. Guo and D. Jasnow,
  {\it J.\ Stat.\ Phys.}\ {\bf 41}, (1985) 353.

\item{[\cite{HuhnDomb}]}W. Huhn and V. Dohm,
  {\it Phys. Rev. Lett.}\ {\bf 61}, (1988) 1368.

\item{[\cite{ChenDohmEsser}]}X.S. Chen, V. Dohm and A. Esser,
  {\it J. de Phys. I France}\ {\bf  5}, (1995) 205.

\item{[\cite{KrechDietrich}]}M.\ Krech and S. Dietrich,
  {\it Phys.\ Rev.}\ {\bf A46}, (1992) 1886;
  {\it Phys. Rev.}\ {\bf A46}, (1992) 1922.

\item{[\cite{FerdFisher}]}A.E.\ Ferdinand and M.E.\ Fisher,
  {\it Phys.\ Rev.} {\bf 185}, (1969) 832.

\item{[\cite{Hertz}]}J. A. Hertz,
  {\it Phys.\ Rev.}\ {\bf B14}, (1976) 1165.

\item{[\cite{BkNiM}]}G.A. Baker, B.G. Nickel and D.I. Meiron,
{\it Phys. Rev.}\ {\bf B17},
(1978) 1365.

\item{[\cite{Expts}]}I.\ Rhee, F.M.\ Gasparini and D.J.\ Bishop,
{\it Phys.\ Rev.\ Lett.}\ {\bf 63}, (1989) 410.
\bye